# Motion Mapping Cognition: A Nondecomposable Primary Process in Human Vision

Zhenping Xie

**Abstract**—Human intelligence seems so mysterious that we have not successfully understood its foundation until now. Here, I want to present a basic cognitive process, motion mapping cognition (MMC), which should be a nondecomposable primary function in human vision. Wherein, I point out that, MMC process can be used to explain most of human visual functions in fundamental, but can not be effectively modelled by traditional visual processing ways including image segmentation, object recognition, object tracking etc. Furthermore, I state that MMC may be looked as an extension of Chen's theory of topological perception on human vision, and seems to be unsolvable using existing intelligent algorithm skills. Finally, along with the requirements of MMC problem, an interesting computational model, quantized topological matching principle can be derived by developing the idea of optimal transport theory. Above results may give us huge inspiration to develop more robust and interpretable machine vision models.

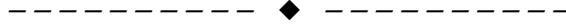

## 1 INTRODUCTION

VISUAL cognition as a core ability of human brain has been highly concerned to explore the mystery of human intelligence. By means of the techniques of neural science and computing science, many explorations [1] and simulations [2] have been deeply performed, and have resulted in huge advancements of machine intelligence [3-5]. Even so, we still cannot say that we have revealed the true secret, or the intrinsic computing mechanism of human visual cognition. Crucially, we still cannot understand how human vision can possess so powerful abilities [6,7].

In this study, I will introduce a novel research view what functions on motion cognition are necessarily required and owned in human vison. And then, the essential computing mechanism of effectively cogniting motion phenomenons of physical world will be further discussed. At last, some very rigorous and interesting conclusions will be put forward, especially I newly introduce a cognitive process "motion mapping cognition" that should be a nondecomposable primary ability in human brain, and could not be effectively modelled in current computing methods. These will give us much inspiration to develop more robust and interpretable machine vision methods.

Based on the researchers' mainstream viewpoints built on brain neural experiments, the human visual cognition (HVC) is a hierarchical process from local visual feature extraction to global object shape recognition [6, 8]. Accordingly, motion tracking was looked as a detection computation of matching similar object shapes in sequential scene images. Above two points have been looked as the defacto "Bible" in computer vision, especially in deep learning idea [2, 9].

However, simple hierarchical process supposition must face two mysteries: 1) how to make the visual feature transforms within different hierarchies be homomorphic and enough robust to the reality of physical scenes [10]; 2) if we believe that evolutional development is the main form of generating human cognition [11], then motion cognition seems to should be a lower function than subtle visual object recognition for surviving demands of animals even human. So, at least monodirectional hierarchical process supposition is inadequate for human visual process. However, how bi-directional hierarchical process works is almost unknown from current research [6].

In this study, I will introduce a novel thought that, there exists a very ground cognitive process, motion mapping cognition (MMC), in human vision. Wherein, I think that dynamic scene change perception is anterior to static visual recognition (including object shape, topology, color, texture, and so on). Obviousely, above consideration is greatly different from existing research, and maybe bring us some extremely novel results.

In above new consideration, motion mapping refers to the mapping matching relations among visual pixels of two sequential frame images. This consideration may be looked as a generalized form of tranditional object shape matching [12].

Moreover, MMC process is completely independent of high-level visual cognitive functions. Wherein, three types of basic mapping relations could be considered for one visual pixel in a previous or subsequent image frame, including: one to one, one to many and one to nil. Here, visual pixels are considered as original image piexels but not feature regions defined in traditional object matching or tracking methods [13, 14].

Along with above discussions, some interesting analy-

---

- *ZP Xie is with the School of Artificial Intelligence and Computer Science, Jiangnan University, Wuxi, Jiangsu, 214122 China.*
  *E-mail: xiezp@jiangnan.edu.cn.*

sis will be given in the rest text. In section 2, some related researches are discussed. In section 3, the detailed definitions and theoretical characteristics are analyzed on motion mapping cognition. Then, modeling challenges are analyzed in section 4, and a very interesting conjecture is presented in section 5. Finally, the main conclusions and contributions are summarized in section 6.

## 2 BACKGROUNDS

### 2.1 Primate visual system

Visual cognition is the most complicated function in human or primate brain. Vast researches based on neurobiological experiments have found some credible mechanisms of micro visual pathways, although the instrinsic principle of how visual system is so powerful in macro scale is still unknown [8, 15].

Nowadys, three kinds of research problems on primate visual systems, as well as their simulations using computers [16], have been exploiting: 1) macroscopical cognitive behaviors or response characteristics to environmental information [11, 17], 2) connectomics and dynamics of visual neural network [8], 3) selective activity rules of individual visual neurons [18, 19]. As a result, following facts are commonly accepted: hierarchical structure, multiple parallel information processing channels, complicated feedbacks from high-level to low-level, learnable/ tunable processing activity.

Ordinarily, neuroscientists attempt to construct the explainable model of primate visual system by hierarchical combination analysis on microscopic behaviors of neurons. In contrast, in order to computably model primate visual system, computer scientists try to subtly understand the working principle of macroscopical cognitive behaviors in accord with the neural activity rules and their connection forms [20].

Obviously, neuroscientists mainly follow the viewpoint of reductionism, while computer scientists primarily believe in the computational possibility of behaviorism. Wherein, the powerful and surprising performance of MuZero [21] and DALL·E2 [22] based on deep neural network and reinforcement learning may be looked as a huge success of computer scientists.

### 2.2 Visual pattern recognition

It is admittedly accepted that, the ability of recognizing different physical objects is one of the basic functions of human brain intelligence. Especially, visual pattern recognition plays the most core role for human life. It can pervieve optical image of physical objects under different environmental conditions, and extract sementic pattern concepts which may almost inerrably coincide to original physical objects. In order to equip this function to intelligent machines, scientists have done vast explorations to uncover and simulate the mystery of human visual recognition. Nowadays, deep neural network (DNN) technologies [2, 9] have become the most powerful tool for above task. Particularly, machine systems can gain better recognition accuracy than human vision in standard image recognition problems [23].

Clearly, DNN is a type of monodirectional signal processing procedure mainly including multilevel feature extraction and feature-based classification. In theory, above procedure may still be viewed as a pattern feature comparing process. Undoubtedly, above way is mainly effective to recognize known pattern objects. Moreover, how to get semantic robustness is still an unsolved key problem for deep neural network [10, 24].

For example, robust visual recognition should require the recognition results satisfy with the semantic composition consistency. That is, the final recognition output and all middle cognition should be completely consistency to a real physical object and its sub-parts. Wherein, interactive selectivity mechanism between neurons of adjacent layers like in human vision may be inevitable. Hence, semantic composition consistency should be a necesssary requirement of machine vision for robustly self-constituting new pattern concepts like in human brain. Obviously, there are no such computational mechanism in current DNN models [10, 24].

According to above analysis, we may also have reason to deem that there may exist other fundamental visual processing ways to support consistency semantic visual cognition in human brain system.

### 2.3 Object tracking and motion perception

In our three-dimensional physical world, the motion of an object is a universal phenomenon. In order to survive effectively, the human brain must accurately and robustly understand the real movement changes of any physical object. Nowadays, the term "object tracking" is considered in machine vision to model the above ability of human brain.

Object tracking methods usually intend to estimate the location updation of the bounding boxes of initialized tracking objects [14, 25]. Concretely, their main process is to find the optimal matching between adjacent regions for two consecutive images. Commonly, optical flow, feature point matching, and region similarity searching methods are three types of mainstream algorithm ideas.

However, from the perspective of visual semantics, human vision should be able to accurately perceive the physical semantic states of object motion in three-dimensional space, including the rotation, translation, deformation of an object, even the separation and combination of multiple objects. Obviously, existing object tracking methods still don't fully cover above abilities, nor do they make some preliminary theoretical discussions on the above issues.

## 3 MOTION MAPPING COGNITION PROBLEM

### 3.1 Basic concepts

In human cognition, how to perceive physical object motion should be a very fundamental ability. Wherein, basic motion forms may be considered as translation, rotation, and deformation. Moreover, I think that most basic ability in motion cognition should be to cognize the point mapping relations between two consecutive frame images containing motion objects. Figure 1 illustrates the map-pling relation examples on three basic motion patterns.

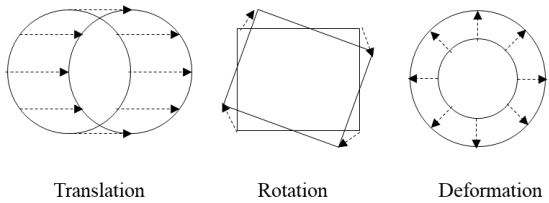

Fig. 1. The mapping relation examples on three basic motion forms.

In figure 1, three kinds of basic motion forms on simple shapes are illustrated including translation, rotation, and deformation. For each demo, there are two rigid shape objects displayed by solid closed lines, and their mappling relations are denoted by dashed straight arrowlines.

For these three mapping examples, we may easily understand their moving process. But for computer, is there a universal computing model that can solve these different motion cases？ Or, if such computing model exists, then what properties it should have？ Furthermore, we may ask: does such universal computing model exist in human cognitive system? To some extent, it is difficult to provide a direct and complete solution of this kind of problem. In this study, I will try to advance some theoretical results.

It seems to be believable that, human even animals can almost seamlessly and accurately understand any physical motion tracks in our survival environment. By simple analysis we may know, three following problems have to be solved to gain the above ability.

1) Cognitive consistency problem for different possible motion tracks.

2) The real-time and subtle requirements of cognitive processing to any motion scenes.

3) Polymorphism problem of motion matching.

For simplicity, I will represent above three prblems as three key functional requirements for human-like motion cognition: *cognitive consistency, real-time sensitivity, and polymorphism*.

The cognitive consistency indicates that the perceived motion situations should keep consistency to true physical motion of observed objects. Real-time sensitivity requires that the computational processing is very efficient and sensitive on two consecutive frame images. The polymorphism denotes that the cognitive motion mapping relations may be many-to-many relationship not only one-to-one relationship on image pixels.

### 3.2 Formal descriptions

According to above discussions, several formal definitions on motion mapping cognition (MMC) are introduced as follows.

*Definition 1*: Motion mapping cognition (MMC) is defined as a process that can seamlessly perceive the accurate motion tracks of three-demensional physical objects by visual information processing. For MMC, it has three core performance requirements: cognitive consistency, real-time sensitivity, and polymorphism.

*Definiiton 2*: The *cognitive consistency* presents that, the output of MMC should be consistency to true physical motion modes.

*Definiiton 3*: The *real-time sensitivity* presents that, MMC can capture the subtle changes of motion modes in real time for any visual scene.

*Definiiton 4*: The *polymorphism* presents that, the outputs of MMC must support many-to-many mapping relations for two consecutive frame images. For some scenes, one object in physical world will split into multiple objects in its moving process, and vice versa.

In order to further explain above concepts, figure 2 and figure 3 containing four motion demos are illustrated. For each demo, two motion stages are respectively labeled by the nubmer 1 and 2 with a circle. And the motion is considered to be starting from stage 1 to stage 2. Similar to fig.1, object shapes are denoted by solid lines, and possible mapping relations are denoted by dashed arrowlines.

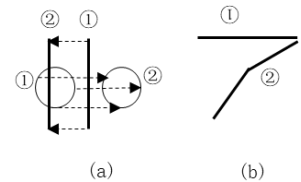

Fig. 2. Two demos on the cognitive consistency problem.

For the secene displayed in fig.2(a), there are two objects, a line segment and a circle. However, two objects move toward two opposite directions respectively. Correspondingly, apparent motion mapping relations are depicted using dashed arrowlines. From our experience, such motion mapping relations can be intuitively cognized. However, for the demo in fig.2(b), there is no exact object match can be easily imaged between two consecutive stage images. Then, what mapping relations will be formed by human visual cognition? Obviously, we cannot directly image what is the most reasonable motion manner. Maybe, the main reason is that, the shape difference between two stage images is bigger than common cases. In real scenes, human usually perceive physical object motions with progressive small changes. Even so, we still can put forward an interesting problem, what mapping perceiton will form by human vision if such two image frames are quckily displayed for human eyes. Its solution form might uncover some mysteries on how human vison can own cognitive consistency on MMC.

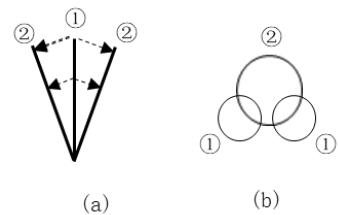

Fig. 3. Two demos on the polymorphic motion problem.

In fig.3(a), a line segment is splitted into two line-segments in two consecutive images. In this case, the mapping relations will contain one-to-two matches, and the most natural motion manner is depicted using dashed arrowlines in the figure. In the fig.3(b), another polymorphic example is exhibited. Similar to fig.2(b), it is difficult to directly depict the most reasonable mapping relations. Different persons may give out different motion tracks that they imaged. Here, I do not concern in the cognitive difference produced by different persons, but focus on the existence of such cognitive ability in human vision. Moreover, such cognitive ability of human being seems to be very robust to our physical world.

According to above further discussions on MMC problem, it should be very difficult to completely clarify the computing mechanism because of huge complexity of real worlds. Next, I will analyze the functional hierarchy of MMC in whole human visual cognition system.

### 3.3 Functional hierarchy of MMC

Here, the functional relations will be analyzed between MMC and existing visual processing tasks including object tracking, object recognition and object extraction etc.

According to the definitions of MMC, it could be known that MMC should be a more fundamental function than object tracking problems. Concretely, the solutions of object tracking problems in existing research do not concern with the complete requirements of MMC problem. In addition, there is also no obvious method that can convert from the solutions of object tracking problems to the solutions of MMC problems. On the contrary, the solutions of MMC problems must contain the solution information of object tracking problems.

For visual object recognition problems, I think they mainly require three functional modules including historic image memorizing, searching, and matching. For MMC problem, we can find that any effective solving algorithm should contain above three functions. Concretely, MMC problem refers to at least two frame images, its solving algorithm must own the ability of memorizing historical image information, and searching possible optimal mapping matching. In brief, if MMC problem could be well resolved by certain algorithms, then visual object recognition tasks also could be well performed using those algorithms in theory.

For object extraction problems, two types of cases should be differently considered including still images and video images. If an object is moving in a video sequence, then its correlation relations among sequential frame images might be resolved by MMC solving algorithms. So, the object could be easily extracted from the image background. When concerning with still images, pattern objects could be firstly recognized by integrating visual object recognition processing. Then, valid object mapping matching between current observation and historical memory can be obtained according to MMC solving algorithms. So, object extraction problems also could be easily solved based on effective MMC solving algorithms.

In addition, other human cognitive abilities including environmental space understanding and individual space self-positioning may tightly relate to visual cognitive processing. Intuitively, if physical objects in sequential video images could be ideally matched by means of MMC solving algorithms, then space cognition might be implemented by integrating personal self-body's motion information and historical experience.

According to above sketchy discussions, we may think that MMC problem may be a very fundamental visual processing requirement for human vision. Its solving algorithms could strongly contribute to effectively solving other multiple visual processing functions. So, I believe that *motion mapping cognition should be a lower-level visual cognitive function comparing to commonly concerned visual functions* including object tracking, object recognition, object extraction, scene space structure understanding etc. Correspondingly, all existing viewpoints on human visual process cannot contribute to effectively solving MMC problem.

Maybe, above analysis seems a little rude, but the deductive conclusion on MMC's level position in whole human visual system should be credible. Undoubtedly, it should be a necessary and more gound visual cognitive ability. Wherein, the most mysterious fact is that, human vision or human cognition seems to own infinite ability of congruously understanding the physical world outside our bodies. Similarly, MMC problem is clearly resolvable in human visual system according to our common sense.

## 4 CHALLENGE OF MODELLING MMC

Despite MMC problem is so important, how to solving it seems to be very trouble. Next, I will give some discussions on existing related methods if they are used to solve MMC problem.

### 4.1 Neural network models

Neural netork (NN) process is now viewed as an elementary computing mechanism in human brain [1]. And artificial neural networks also have making great success in many fields of AI [9]. As we have known, the main computing ways of NN process include multiple channel signal selection, hierarchical convergent computing, inter-layer parameter inter-selection, and neural parameter' learnable updation [6,8].

For the requirement of real sensitivity in MMC, hierarchical convergent computing process might be effective. By using deep neural network model, real-time scene recognition on object motion changes can be performed. Crucially, deep hierarchical computing structure may support fast forward signal processing from original inputs to final recognition outputs.

For the requirement of polymorphism in MMC, we could consider that, different mapping relations may be represented by different hierarchical channels. Despite few researches is reported, above considerations may be reasonable and realizable in theory.

Finally, for the requirement of cognitive consistency, if only neural parameters are trained to model given finite samples or labeled examples, then it is impossible for trained neural networks to model all possible physical motion mapping relations with rigorous cognitive consistency. Above problem is also the most deficiency of current deep neural networks. In addition, we still can not make trained deep neural network be explainable and engouh robust, which also limits the possibility of solving MMC problem using existing neural network models.

### 4.2 Graph matching algorithms

Graph matching algorithms mainly try to obtain matched mapping between two graph structures. They are useful computational tools for many visual tasks [26, 27].

For existing graph matching algorithms, graph feature extraction and feature matching are two main algorithm modules. Graph feature extraction compute feature vectors for each graph node or sub-graph by constructing feature functions on topological graphs [26]. And feature matching processing usually uses optimization techniques to obtain point-to-point or region-to-region matching. Obviously, above processing could not model polymorphic mapping relations. In addition, it is completely no way for existing optimization algorithms to guarantee that estimated mapping relations are optimally consistent to real physical motions.

In a word, current graph matching algorithms seem unavailable for solving MMC problem.

### 4.3 Optimal transport theory

Concerning with the characteristics of MMC, the optimal transport theory (OTT) [28, 29] may be considered as another idea origin of possible solving algorithms. OTT is a very powerful tool to get the optimal mapping relation between two given signal distributions. Its standard form can be written as,

$$\arg\min_{T} OTT(\mu,\nu) = \int_X c(x,T(x))d\mu(x) \quad (1)$$
$$wherein \quad T(x): \mu \to \nu$$

The above equation is also called as minimum Wasserstein distance optimization equation. Wherein, $c(x,T(x))$ denotes a transport cost function, it may be Euclidean distance in normal physical scenes.

For OTT, it usually requires the following constraint

$$\int_X d\mu(x) = \int_Y d\nu(y) \quad (2)$$

Wherein, X and Y are two set spaces equipped with measures $\mu$ and $\nu$ respectively. Although OTT has shown its powerful ability in some important problems such as the explanation on training generative adversarial networks (GAN) [30], howerver, if the objective form of OTT is employed to model MMC problem, three following troubles seem to be unsolvable.

(1) For an object that is largening or shrinking at its boundary region, its two consecutive frame images will present some difference only on its boundary region. If we directly use OTT to model above motion mapping, then possible solutions will require that every tiny region has at least a small displacement because of the requirement of constraint equation (2). Obviously, above solution form is not consistent to true physical motion or the cognitive result of human vision. (**Boundary shaping problem**)

(2) For a scene in which there are a still object and a moving point light scource, then its two consecutive motion frame images will present some changes on illumination intensity for most same regions. So, the mapping relations between two consecutive images will produce jitter according to the optimal solutions of OTT. Such solutions are clearly unreasonable and inaccurate. (**Illumination invariance problem**)

(3) The Mather's shortening lemma is a basic feature for the solutions of OTT. However, this limitation may be improper for cognizing rollover motion in MMC problem, because that the motion mapping of roolover motion for a closed object will necessarily contain intersecting mapping relation lines. (**Rollover problem**)

According to above disscusions, the objective of OTT might not be directly available for MMC problem. Even so, the solution form of OTT should be the most relevant exploration on mapping cognition between two geometrical objects.

### 4.4 The theory of topological perception

According to existing researches, the theory of topological perception (TTP) presented by Lin Chen [31,32] should be the most powerful supposition to interpret MMC problem. Specifically, Chen's theory may explain the phenomenons on cognitive consistency and real-time sensitivity in MMC problem.

Nevertheless, the polymorphism mechanism in MMC problem is still not be uncovered by Chen's TTP. Besides, Chen's theory only gives some experimental and qualitative explanations on the topological property of visual perception phenomenons, but no further computing mechanism research. That is, there is no concrete exploitation on what computational process may be feasible for TTP problem. So, MMC problem may be looked as a very meaningful development of Chen's TTP.

## 5 AN INTERESTING CONJECTURE

As discussed above, there is no existing computational approach that can successfully model the computing requirements of MMC problem. However, we may believe that, human bain can almost perfectly solve MMC problem. So, what mechanism might be feasible and necessary to solve MMC problem?

Broadly speaking, MMC problem also could be looked as a kind of generalized graph matching problem in theory. Wherein, the matching or mapping objective should be very different from existing algorithm considerations. So, more idea inspirations and explorations might be introduced, such as topological data analysis [33] and quantum machine learning [34, 35] methods.

By rethinking above algorithm ideas, especially the relations between the OTT idea and the requirements of MMC problem, a new learning mechanism could be inferred as follows.

$$\arg\max_{\varphi} QTM(\varphi) = \\ \int_{X,Y} \frac{\kappa(\varphi,x,y,\mathrm{M})}{c(x,y)+\kappa(\varphi,x,y,\mathrm{M})} \varphi^{xy}\varphi^{yx} dxdy \quad (3)$$

Wherein, $X,Y \subset R^2$ are the space regions with respect to two consecutive images, and $\varphi$ is a bi-directional mapping function between $X$ and $Y$. Here, $\varphi$ could be looked as a more flexible form than $T(x)$ in OTT, and $c(x,y)$ also denotes a transport cost function like in OTT. In equation (3), $\kappa(\varphi,x,y,\mathrm{M})>0$ is a priori scale function with hyperpameter $\mathrm{M}$, which is internally required in mathematical form containing priori experience or knowledge informaiton.

In addition, following constraints are considered in equation (3) for given images A and B on $X$ and $Y$ respectively,

$$\sum_y (\varphi^{xy})^2 = A_x \quad (4)$$

$$\sum_x (\varphi^{yx})^2 = B_y \quad (5)$$

It should be pointed out that, $\varphi^{xy}$ and $\varphi^{yx}$ are all nonnegative and might be unequal.

For above learning mechanism according to equation (3) with constraints (4) and (5), I call it as **quantized topological matching (QTM) principle.**

Like the description of QTM, OTT could be rewritten as follows.

$$\arg\min_{\varphi} OTT(\varphi) = \int_{X,Y} c(x,y) \varphi^{xy} \varphi^{yx} dxdy \quad (6)$$

Wherein, a bidirectional symmetry condition $\varphi^{yx}=\varphi^{xy}$ and same constraints of (4) and (5) are required.

By comparing the objective forms (3) and (6), we may say that they are dual forms except for different constraint requirements. Wherein, some more restrict constraints must be considered in OTT on account of its basic principle. Although more flexible $\varphi^{xy}$ and $\varphi^{yx}$ could be allowed in QTM, but an extra scale function $\kappa(\varphi,x,y)$ must be introduced for mathematical rationality. Fruthermore, how to deal with $\kappa(\varphi,x,y,\mathrm{M})$ seems to be very trouble, and so how to solving QTM equation has to be studied in further long-term works.

Although the core principle is some different between the objectives of QTM and OTT, but they all concide to the following requirements of MMC problem.

1) Sensitivity to tiny motion change.
2) Mapping polymorphism for topological coalition and separation.

However, compared to OTT, the QTM objective might be supportable for the specific requirements of MMC including, boundary shaping problem, illumination invariance problem, and rollover problem.

Although we can not confirm whether QTM will be completely effective or approximatively feasible for solving MMC problem, but it could be believed that the QTM principle will give us very meaningful inspiration for solving MMC problem, and will provide huge enlightenment to develop interpretable and robust machine vision algorithms like in human vision.

## 6 CONCLUSIONS

In order to explpoit the mystery of human cognition, a completely new fundamental problem in human vision, motion mapping cognition (MMC) is presented. As an inevitable cognitive process, MMC also may be very effective for logically explaining most cognitive functions of human vision, if MMC problem could be reasonably solved in theory.

In this study, I further indicate the reasonability of Chen's theory of topological perception, and clarify that MMC process as an extension of Chen's theory should be a nondecomposable basic visual cognitive process in human vision.

In addition, a quantized topological matching (QTM) principle was deduced according to the requirements of MMC problem and the ideas of optimal transport theory. Clearly, QTM principle may provide us great idea inspiration to exploit more robust and interpretable visual processing models like in human brain.

## ACKNOWLEDGMENT

This work was supported by the grants from National Natural Science Foundation of China (No.62272201, 61872166).

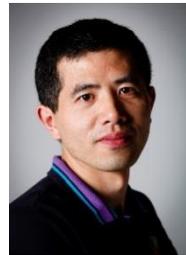
**Zhenping Xie** received the B.Eng. and Ph.D. degrees from Jiangnan University, Wuxi, China, in 2002 and 2008, respectively. He is currently a professor with the School of Artificial Intelligence and Computer Science, Jiangnan University. He has authored or co-authored of about 50 research papers in academic journals or conferences. His research interests include knowledge computing and cognitive learning.